\documentstyle[fleqn]{article}
\begin{document}
\LARGE
\begin{center}
Mutually Unbiased Bases and Finite Projective Planes
\end{center}
\vspace*{-.1cm} \Large
\begin{center}
Metod Saniga,$^{1}$ Michel Planat$^{2}$ and Haret Rosu$^{3}$

\vspace*{.15cm} \small $^{1}$Astronomical Institute, Slovak
Academy of
Sciences, 05960 Tatransk\' a Lomnica, Slovak Republic\\
$^{2}$Institut FEMTO-ST, D\'epartement LPMO, CNRS,
32 Avenue de l'Observatoire, 25044 Besan\c{c}on, France\\
$^{3}$Department of Applied Mathematics, IPICyT, Apdo Postal 3-74
Tangamanga, San Luis Potos\'{\i}, Mexico
\end{center}

\vspace*{-.4cm} \noindent \hrulefill

\vspace*{.2cm} \small \noindent
{\bf Abstract}\\

\noindent It is conjectured that the question of the existence of
a set of $d+1$ mutually unbiased bases in a $d$-dimensional
Hilbert space if $d$ differs from a power of prime is intimatelly
linked with the problem whether there exist projective planes
whose order $d$ is not a power of prime.

\noindent \hrulefill

\vspace*{.5cm} \normalsize \noindent Recently, there has been a
considerable resurgence of interest in the concept of the
so-called mutually unbiased bases [see, e.g., 1--7], especially in
the context of quantum state determination, cryptography, quantum
information theory and the King's problem. We recall that two
different orthonormal bases $A$ and $B$ of a $d$-dimensional
Hilbert space $\cal{H}$$^{d}$ are called {\it mutually unbiased}
if and only if $|\langle a|b\rangle|= 1/\sqrt{d}$ for all
$a$$\in$$A$ and all $b$$\in$$B$. An aggregate of mutually unbiased
bases is a set of orthonormal bases which are pairwise mutually
unbiased. It has been found that the maximum number of such bases
cannot be greater than $d+1$ [8,9]. It is also known that this
limit is reached if $d$ is a power of prime. Yet, a still
unanswered question is if there are non-prime-power values of $d$
for which this bound is attained. The purpose of this short note
is to draw the reader's attention to the fact that the answer to
this question may well be related with the (non-)existence of
finite projective planes of certain orders.

A finite {\it projective} plane is an incidence structure
consisting of points and lines such that any two points lie on
just one line, any two lines pass through just one point, and
there exist four points, no three of them on a line [10]. From
these properties it readily follows that for any finite projective
plane there exists an integer $d$ with the properties that any
line contains exactly $d+1$ points, any point is the meet of
exactly $d+1$ lines, and the number of points is the same as the
number of lines, namely $d^{2}+d+1$. This integer $d$ is called
the {\it order} of the projective plane. The most striking issue
here is that the order of known finite projective planes is a
power of prime [10]. The question of which other integers occur as
orders of finite projective planes remains one of the most
challenging problems of contemporary mathematics. The only
``no-go" theorem known so far in this respect is the Bruck-Ryser
theorem [11] saying that there is no projective plane of order $d$
if $d-1$ or $d-2$ is divisible by 4 and $d$ is not the sum of two
squares. Out of the first few non-prime-power numbers, this
theorem rules out finite projective planes of order 6, 14, 21, 22,
30 and 33. Moreover, using massive computer calculations, it was
proved by Lam [12] that there is no projective plane of order ten.
It is surmised that the order of {\it any} projective plane is a
power of a prime.

>From what has already been said it is quite tempting to
hypothesize that the above described two problems are nothing but
different aspects of one and the same problem. That is, we
conjecture that {\it non-existence of a projective plane of the
given order $d$ implies that there are less than $d+1$ mutually
unbiased bases (MUBs) in the corresponding $\cal{H}$$^{d}$}, and
vice versa. Or, slightly rephrased, we say that if the dimension
$d$ of Hilbert space is such that the maximum of MUBs is less than
$d+1$, then there does not exist any projective plane of this
particular order $d$.

An important observation speaking in favour of our claim is the
following one. Let us find the minimum number of different
measurements we need to determine uniquely the state of an
ensemble of identical $d$-state systems. The density matrix of
such an ensemble, being Hermitian and of unit trace, is specified
by $(2d^{2}/2) - 1 = d^{2}-1$ real parameters. As a given
non-degenerate measurement applied to a sub-ensemble gives $d-1$
real numbers (the probabilities of all but one of the $d$ possible
outcomes), the minimum number of different measurements needed to
determine the state uniquely is $(d^{2}-1)/(d-1)= d+1$ [8]. On the
other hand, it is a well-known fact [see, e.g., 13] that the
number of $k$-dimensional linear subspaces of the $n$-dimensional
projective space over Galois fields of order $d$ is given by
\begin{eqnarray}
\left[
\begin{array}{c}
n+1 \\
k+1 \end{array}
 \right]_{d} \equiv
\frac{(d^{n+1}-1)(d^{n+1}-d)...(d^{n+1}-d^{k})}
{(d^{k+1}-1)(d^{k+1}-d)...(d^{k+1}-d^{k})}, \nonumber
\end{eqnarray}
which for the number of {\it points} ($k$=0) of a projective {\it
line} ($n$=1) yields $\left[
\begin{array}{c}
2 \\
1 \end{array}
 \right]_{d} = (d^{2}-1)/(d-1)= d+1$.

Another piece of support for our conjecture comes from the ever
increasing use of geometry in describing simple quantum mechanical
systems. Here we would like to point out the crucial role the
so-called Hopf fibrations play in modelling one-qubit, two-qubit
and three-qubit states. Namely, the $s$-qubit states, $s=1,2,3$,
are intimately connected with the Hopf fibration of type
$S^{2^{(s+1)}-1} \stackrel{S^{2^{s}-1}} \longrightarrow S^{2^{s}}$
[14--16], and there exists an isomorphism between the sphere
$S^{2^{s}}$, $s=1,2,3$, and the {\it projective line} over the
algebra of complex numbers, quaternions and octonions,
respectively [17].

Perhaps the most serious backing of our surmise is found in a
recent paper by Wootters [18]. Associating a line in a finite
geometry with a pure state in the quantum problem, the author
shows that a complete set of MUBs is, in some respects, analogous
to a finite {\it affine} plane, and another kind of quantum
measurement, the so-called symmetric informationally complete
positive-operator-valued measure (SIC POVM), is also analogous to
the same configuration, but with the swapped roles of points and
lines. It represents no difficulty to show that this ``dual" view
of quantum measurement is deeply rooted in our conjecture. To this
end, it suffices to recall two facts [10].  First, any affine
plane is a particular subplane (subgeometry) of a projective
plane, viz. a plane which arises from the latter if one {\it
line}, the so-called ``line at infinity," is deleted. Second, in a
projective plane, there is a perfect {\it duality} between points
and lines; that means, to every projective plane, $S_{2}$, there
exists a dual projective plane, $\Sigma_{2}$, whose points are the
lines of $S_{2}$ and whose lines are the points of $S_{2}$ [19].
So, {\it affinizing} $S_{2}$ means deleting a {\it point} of
$\Sigma_{2}$ and thus, in light of our conjecture, qualitatively
recovering the results of Wootters, shedding also important light
on some other of the most recent findings [20,21]. The latter
reference, in fact, gives several strong arguments that there are
no more than three MUBs in dimension six, the smallest
non-prime-power dimension.

Finally, at the level of applications, finite projective spaces
have already found their proper place in classical enciphering
[10]. By identifying the points of a (finite) projective space
with the eigenvectors of the MUBs endowed with a Singer cycle
structure one should, in principle, be able to engineer quantum
enciphering procedures. These should play a role in the emerging
quantum technologies of quantum cryptography and quantum computing
[22].

\end{document}